\documentclass[usenatbib]{mnras}

\usepackage{newtxtext,newtxmath}

\usepackage[T1]{fontenc}

\DeclareRobustCommand{\VAN}[3]{#2}
\let\VANthebibliography\thebibliography
\def\thebibliography{\DeclareRobustCommand{\VAN}[3]{##3}\VANthebibliography}

\usepackage{graphicx}	
\usepackage{bm}
\newcommand{\sref}[1]{Section~\ref{#1}}
\newcommand{\fref}[1]{Figure~\ref{#1}}

\newcommand{\kms}[0]{km~s^{-1}}

\title[Improved statistic for lensed GW events]{Improved statistic to identify strongly lensed gravitational wave events}

\author[More \& More]{
Anupreeta More,$^{1,2}$\thanks{E-mail: anupreeta@iucaa.in}
and Surhud More,$^{1,2}$
\\
$^{1}$The Inter-University Centre for Astronomy and Astrophysics (IUCAA), 
Post Bag 4, Ganeshkhind, Pune 411007, India
\\
$^{2}$Kavli Institute for the Physics and Mathematics of the Universe (IPMU), 5-1-5 Kashiwanoha, Kashiwa-shi, Chiba 277-8583, Japan
}

\date{Accepted XXX. Received YYY; in original form ZZZ}

\pubyear{2021}

\begin{document}
\label{firstpage}
\pagerange{\pageref{firstpage}--\pageref{lastpage}}
\maketitle

\begin{abstract}
As the number of detected gravitational wave sources increase with increasing sensitivity of the gravitational wave observatories, observing strongly lensed pair of events will become a real possibility. Lensed GW events will have very accurately measured time delays, as well as magnification ratios. If the corresponding lens system can be identified electromagnetically and the redshifts of the lens and the host galaxy can be measured, such events can be used to constrain important astrophysical parameters of the lens system. As most lensing events will have image separations that will be significantly smaller than the GW event localization uncertainties, it is important that we develop diagnostics that will aid in the robust identification of such lensed events. We define a new statistic based on the joint probability of lensing observables that can be used to discriminate lensed pair of events from the unlensed ones. To this end, we carry out simulations of lensed GW events to infer the distribution of the relative time delays, relative magnifications sub-divided by the type of lensed images. We compare this distribution to a similar one obtained for random unlensed event pairs. Our improved statistic can be used to improve the existing ranking approach adopted by the search pipelines in order to down-select event pairs for joint parameter estimates. The distributions we obtain can further be used to define more informative priors in joint parameter estimation analyses for candidate lensed events.
\end{abstract}




\section{Introduction}

Gravitational waves (GWs) from compact binary mergers, comprising of black holes and/or neutron stars, are routinely being discovered now by gravitational wave observatories such as the Advanced LIGO \citep[][]{Aasi2015} and Advanced Virgo \citep{Acernese2015}. Over 50 GW events are reported till date \citep[][]{Abbott_2019, Venumadhav2020, Abbott_2021}. The rates are expected to increase substantially with the upgrades in the sensitivity of Advanced LIGO  complemented by additional detectors such as KAGRA \citep[][]{akutsu2020overview} which is already up and running  and LIGO-India \citep{Abbott_2020} which is expected to turn on in the near future.  With the next generation observatories such as the Cosmic Explorer \citep[e.g.][]{AbbottCE_2017} and Einstein Telescope \citep[][]{Hild2011, Punturo2014} being planned, the sensitivity of gravitational wave detection, and hence our ability to use them as an investigative tool to probe the Universe will increase tremendously. One such example, is the gravitational lensing of GWs \citep[e.g.][]{Wang1996, Nakamura1998}, which was predicted even before the discovery of GWs. Lensed GW events are expected to be found and studied with more sensitive detectors \citep[e.g.,][]{Li2018, Oguri2018, Yang2021}.

Detection of gravitational lensing of GW events will confirm one of the important predictions of general relativity entirely from gravitational waves. The magnification provided by gravitational lensing could help detect GW events from higher redshifts, which would otherwise have remained sub-threshold. The discovery of such lensed GW events could have important implications on the rate evolution of BBH mergers \citep[e.g.][]{Abbott2021_lens} or their formation channels \citep[e.g.][]{Mukherjee2021}. In addition the lensed events could be used to obtain interesting astrophysical information about the lensing galaxy. If the gravitational lens and the potential host galaxy of the GW event can be identified along with an electromagnetic counterpart for events such as NS$-$BH mergers, then they could be further targeted electromagnetically to obtain redshifts for the lens and the source GW event. This information when combined with the observable time delays and magnification ratios could even be useful to constrain cosmological parameters such as the Hubble constant \citep[e.g.][]{Liao2017}.

If GW events get strongly lensed by an intervening massive deflector such as a galaxy or a galaxy cluster, the GWs emitted by the coalescing binaries from the distant Universe will get deflected in multiple directions and arrive at the observer with a modification to the waveform expected in the absence of gravitational lensing. In addition to a time delay, these events will appear to have identical intrinsic parameters (separated only by statistical noise), such as the redshifted masses of their component binaries. However the waveforms will undergo a change in amplitude due to the lensing magnification, in addition to changes in its shape depending upon the path the waves take along their journey to the observer.

Many searches for lensing signatures have been carried out on the GW data obtained from the multiple LIGO--Virgo observing runs \citep[e.g.,][]{Haris2018, Hannuksela2019, Dai2020Morse, Liu2021, Abbott2021_lens} which demanded development of new algorithms and techniques. The identification of lensed events is a model selection problem and requires evaluation of the odds of the lensing hypothesis compared to the random event pairs hypothesis. One such technique relies on the consistency of the posterior distribution of the sky localization and intrinsic parameters unaffected by lensing for two candidate lensing events. The posterior overlap search is applied to the GW events detected by the detection pipelines and whose intrinsic parameters have been inferred. If the posteriors for a pair of GW events are consistent then this may suggest that the likelihood of lensing is higher than such events being produced at random \citep{Haris2018}.

Furthermore, hierarchical Bayesian framework has been developed to identify strongly lensed events by considering the gravitational lensing scenario as well various GW selection effects \citep[][]{Lo2021}. Here joint posterior distributions of intrinsic parameters as well as the lensing parameters (such as the relative magnifications, time delays and image types) are obtained for any given pair of GW events. Search for lensed counterparts in the sub-threshold GW candidate event catalogs have also been attempted as some of the multiply lensed events may have lower relative magnifications or be de-magnified to such an extent that they lay buried among the low-confidence triggers rather than the confirmed super-threshold GW events \citep[e.g.,][]{Li2019, McIsaac2020}. 

The relative time delays and magnifications between the pair of events are the principle observables in gravitational lensing. Because gravitational wave detectors measure waveform amplitudes instead of intensities, they can additionally measure phase shifts between the candidate event pairs \citep{Dai2017} and thus help to determine the possible image types (see Section~\ref{sec:gl}). In previous lens searches, the expected distribution of time delays, that of lensing magnification, or of the phase differences, have been used independently as priors to analyse selected pairs of events \citep[e.g.,][]{Haris2018, Hannuksela2019, Dai2020Morse, Liu2021, Lo2021}. In this work, we obtain the joint distribution of these three lensing observables and propose to use it to help in the identification of candidate lensed events and their analysis. We compare it to the distributions of similarly constructed observables from random unlensed events. This distribution can be used for both a robust preselection of candidate lensed GW events and serve as useful priors in the joint parameter estimation (PE) studies \citep{Lo2021}.

The structure of the paper is as follows. In \sref{sec:gl}, we briefly go through the principles of gravitational lensing and introduce the various terms we will use in the rest of the paper. In \sref{sec:bayes}, we present a framework to use the joint distribution of lensing observables for the preselection of candidate events. In \sref{sec:simgw}, we present the framework for generating the simulated lensed and unlensed GW events. In \sref{sec:statprop}, we show the resulting joint distribution of the different lensing observables. We summarize our results in \sref{sec:summary}.

\section{Gravitational lensing}
\label{sec:gl}


In this section, we introduce the basics of strong gravitational lensing \citep[see e.g. ][for the theory of gravitational lensing]{Schneider1992}. In strong gravitational lensing, multiple images of a distant background source are formed due to the gravitational potential of an intervening massive object such as a galaxy or a cluster of galaxy.

The lensing properties of the system can be expressed in terms of a dimensionless scalar field called the Fermat potential which is proportional to the time of arrival of photons from a given true position of the source in the plane of the sky ($\beta$) and the observed position of its corresponding image ($\theta$). The Fermat potential is given by
\begin{equation}
    \phi(\bm{\theta,\beta}) = \frac{1}{2}(\bm{\theta}-\bm{\beta})^2 - \psi(\bm{\theta})) \,,
    \label{eq:fermpot}
\end{equation}
where the first term corresponds to a geometric term while the second term corresponds to the gravitational contribution of the intervening object. The geometric term consists of the difference between the angular positions of the images ($\bm{\theta}$) and that of the source ($\bm{\beta}$) whereas the lensing potential ($\psi$) in the second term is related to the lens mass density profile $\Sigma(\bm{\theta})$ as follows
\begin{equation}
\psi(\bm{\theta})= \int \rm {d}^2\bm{\theta'} \kappa(\bm{\theta'}) \rm{ln}|\bm{\theta}-\bm{\theta'}|\,.
    \label{eq:lenspot}
\end{equation}
Here, the quantity $\kappa(\bm{\theta}) = \Sigma(\bm{\theta})/\Sigma_{\rm{cr}} $ is the dimensionless convergence and the critical surface density { $\displaystyle \Sigma_{\rm{cr}}= \frac{c^2}{4\pi G} \frac{ D_{\rm s}}{ D_{\rm d} D_{\rm ds}} $}, is defined using the angular diameter distances to the source ($D_{\rm s}$), the lens ($D_{\rm d}$) and between the lens and the source ($D_{\rm ds}$).

According to Fermat's principle, the lensed images are formed at the stationary points in the Fermat's time delay surface \citep[see e.g.][for details]{Blandford1986}. There are three types of stationary points on the time delay surface, namely, a minimum, a saddle and a maximum. The lensed images formed at these points are referred to as Type I, Type II or Type III images, respectively. 

The relative time delay for the lensed images at $\bm{\theta_1}$ and $\bm{\theta_2}$ can be written as
\begin{equation}
    \Delta t = \frac{(1+z_{\rm l}) D_{\rm d} D_{\rm s}}{c  D_{\rm ds}} \left[ \phi(\bm{\theta_1,\beta}) - \phi(\bm{\theta_2,\beta}) \right]
\end{equation}

The relation between the source and the image position can be found by setting $\nabla_{\bm{\theta}}(\phi)= 0 $, and is given by the lens equation,
\begin{equation}
    \bm{\beta} 
    = \bm{\theta}-\nabla_{\bm{\theta}}\psi(\bm{\theta}) = \bm{\theta}-\bm{\alpha}(\bm{\theta}),
    \label{eq:lenseq}
\end{equation}
where the deflection angle, $\bm{\alpha}$ is the gradient of the lensing potential $\psi$, which causes the above equation to be non-linear. Therefore this equation can potentially have multiple solutions $\bm{\theta}$ for a given value of $\bm{\beta}$, which correspond to the multiple lensed images.

The non-linear mapping between the source position and the image position results in a distortion of photon bundles arriving from the source. This distortion can be expressed with the help of the Jacobian $\mathbb{A}_{ij}\equiv[\partial \beta_i/\partial \theta_j]$,
\begin{equation}
\mathbb{A}_{ij} = 
\begin{pmatrix}
1-\kappa-\gamma_1 & -\gamma_2 \\
-\gamma_2 & 1-\kappa+\gamma_1 
\end{pmatrix}
\end{equation}
where the convergence $\kappa=\psi_{11}+\psi_{22}$ is the divergence of the lensing potential, and the two components of the shear are given by $2\gamma_1=(\psi_{11}-\psi_{22})$ and $\gamma_2=\psi_{12}$. The convergence leads to a uniform scaling of any bundles of photons arriving from the source, while the shear corresponds to an anisotropic distortion of the bundles of these photons. At the position of a minimum both the eigenvectors of this matrix are positive, while at a maximum both the eigenvectors are negative. In both cases, this causes the images to have the same parity as the original source. However at a saddle point, the two eigenvectors have opposite signs, which causes a reversal in the parity of the images. The Morse index $n$ is given by half the number of negative eigenvalues of this matrix and is useful in the case of wave propagation as we describe later. Thus, for images that form at the minima of the Fermat potential $n_{\rm Type I}=0$, those that form at a saddle point $n_{\rm Type II}=1/2$ while those images that form at maxima will have $n_{\rm Type III}=1$.

The scalar magnification is given by the inverse of the determinant of the Jacobian (matrix),
\begin{equation}
    \mu\equiv |\det\mathbb{A}|^{-1} = \big[\left(1-\kappa\right)^2-\gamma^2\big]^{-1},
    \label{eq:lensmag}
\end{equation}
where $\gamma=[\gamma_1^2+\gamma_2^2]^{1/2}$ corresponds to the amplitude of the shear at the positions of the images.

The effects of gravitational lensing on gravitational waves can be determined by considering the propagation of the gravitational radiation from the source to the observer using Kirchoff's diffraction integral. In the geometric optics limit, the main contribution to the diffraction integral arise by trajectories taken by the waves near the stationary points of the time delay surface. In the vicinity of the stationary points, the integral reduces to Fresnel integrals and the resultant complex waveform that describes the strain for a given lensed image is the convolution of the original waveform with
\begin{equation}
    V(f) = |\mu_{j}|^{1/2}\exp\left( i2\pi f T_j - i \pi {\rm sgn}[f]n_j \right)
\end{equation}
where $f$ is the observed wave frequency, $T_j$ is the light travel time for image $j$, $\mu_j$ is the signed magnification, $n_j$ is the Morse index defined in the section above for each type of image, and the ${\rm sgn}$ function denotes the sign of the frequency $f$. In addition to the time delay, the waveform gets modified by the amplitude $\mu_j^{1/2}$, and depending upon the image type, also acquires Morse phase shifts given by $n_j \pi$. The waveforms of type I images thus just shift in time and those of Type III undergo a change in sign compared to the original waveform in addition to the shift due to time delays. However, for Type II images, the positive (negative) frequency components experience an additional phase shift of $\pi/2$ ($-\pi/2$). These effects on the waveform are detectable and are used when analyzing pairs of events for joint parameter estimates \citep{Lo2021}.


We note that in realistic scenarios where a lensing galaxy is the deflector for a distant GW source, we may have to worry about further additional fluctuations in the lensing magnification owing to what is conventionally referred to as ``Microlensing'' in the electromagnetic domain studies. In such a case, additional contribution from compact stars, stellar remnants or primordial black holes located within the main deflector may have to be accounted for \citep[e.g.][]{Diego2019,Diego2020,Mishra2021}. 
However, as pointed out in \citet[][]{Mishra2021}, such microlensing effects become significant only when the lensing magnification by the main deflector  reaches moderate-to-high values ($\mu~\gtrsim~15$) which is expected in small fraction of the lensed events. Thus, we ignore any microlensing fluctuations in our analysis.   


\section{Framework to preselect candidate lensed event pairs}
\label{sec:bayes}
Multiple strategies have been implemented to search for gravitational lensing events in the gravitational wave data. The simplest of searches utilizes the magnification induced by gravitational lensing. The lensing magnification, $\mu$ results in a larger amplitude for the GWs, which causes the inferred luminosity distance to be smaller by a factor $\sqrt{\mu}$. If the event is analyzed without taking lensing into account, this will bias the inferred masses of the events to be larger. There have been a handful of GW events detected with unusually high masses for their progenitors and such events have been considered as candidates for gravitational lensing \citep[][]{Abbott2021_lens}. However, our knowledge of the intrinsic population of binary black holes and binary neutron stars is quite uncertain, and it remains unclear whether such events represent an unusual intrinsic population of black holes and neutron stars or are a population affected by gravitational lensing. 
Given that the gravitational lensing rates are expected to be small, the latter scenario thus requires a stronger evidence than just the unusual nature of these events. 

Robust identification of multiple images in the GW data could provide such an evidence. Gravitational lensing will result in a modification of the amplitude of the GW signal as well as the time of arrival of these events at the GW detectors. In addition, lensing can result in a frequency independent phase shift depending upon the type of the image (see \sref{sec:gl}).

Whereas the lensing amplitude differences could be  
adjusted in the inferred distances, the time of arrival in  each pair of events has to be separately analyzed under the lensing hypothesis, in order to provide a quantitative 
comparison of the lensed vs the unlensed scenario.

Therefore, the probability of the lensing hypothesis 
depends upon the overlap of the sky localization of the two events and the parameters like the redshifted component masses of the BBH, dimensionless spin magnitudes, cosine of the spin tilt angles, the cosine of the orbital inclination angle, that do not get affected by lensing. 
Currently, in the lensing search pipelines,  this overlap is computed using the ranking statistic ${\sc B}^{\rm overlap}$ given by,
\begin{equation}
    {\sc B}^{\rm overlap} = \int d\Theta \frac{P(\Theta|d_1)P(\Theta|d_2)}{P(\Theta)}
\end{equation}
where $P(\Theta)$ denotes the priors on these parameters, while $P(\Theta|d_i)$ denotes the posterior distribution of these parameters given the signal from the $i$-th event of the pair.

The above ranking statistic quantifies the overlap of the parameters of the two events except those that get affected by gravitational lensing which are i) the phase of coalescence (affected by parity differences) ii) the time of arrivals (affected by time delays), and iii) the distances (affected by magnification). 

The lensing phase differences $\Delta\varphi$ are expected to have either of these fixed values - $0, \pi/2, 3\pi/2$ or $\pi$ \citep[e.g.][]{Dai2017}. 
The distribution of time delays of the lensed events is significantly different from that of the unlensed events. This is quantified as a ratio of probabilities of the time delays under the lensed and the unlensed hypothesis \citep[see][]{Haris2018},
\begin{equation}
    {\cal R}^{\rm gal} = \frac{P(\Delta t| {\cal H}_{\rm SL})}{P(\Delta t| {\cal H}_{\rm UL})}
\end{equation}

We propose to use a new statistic that accounts for the phase difference, time delay as well as the magnification ratio differences by computing,
\begin{equation}
    {\cal M}^{\rm gal} = \frac{P(\Delta t, \mu_{\rm r}, {\Delta\varphi}| {\cal H}_{\rm SL})}{P(\Delta t, \mu_{\rm r}, {\Delta\varphi}| {\cal H}_{\rm UL})}\,.
    \label{eq:mgal}
\end{equation}
The time delays are measurable with quite exquisite precision in the case of GW events. The determinations of the magnification ratios via their inferred distances, as well as lensing phase shifts can relatively have higher uncertainties. Nevertheless, the new statistic should use more information than that contained in the statistic just based on time delays and is thus expected to perform better as detector sensitivities improve.

Next, we carry out simulations of lensed and unlensed GW events in order to determine the probability distributions of the $ {\cal H}_{\rm L}$ and $ {\cal H}_{\rm UL}$ hypotheses.

\section{Simulations of lensed and unlensed GW events}
\label{sec:simgw}
In this section, we describe the method we use in order to simulate lensed GW events. In short, after drawing from a realistic lens and source BBH populations which are expected to form lens systems, we randomly distribute a large number of BBH sources behind every lens within regions in the source plane which are expected to result in multiple images. Some of these sources get doubly or quadruply lensed. We extract the lensed image properties and retain only such lens systems for which the second brightest lensed image has a lensed network SNR above the threshold of 8\footnote{We also have independently extracted properties of lensed images in case only one of the events is brighter than the network SNR threshold to aid in searches of counterpart images below the threshold.}.

\subsection{BBH source population}
\label{sec:bbh_src}
We consider binary black hole mergers (BBHs) as our background sources and massive early type galaxies as our lens galaxies. For the BBH sources, we use a fit to the population I/II star merger rate densities, $R(z)$, given by eq.~13 in \citep[][]{Oguri2018} 
\begin{equation}
    R(z) = R_0\frac{a_1e^{a_2z}}{e^{a_3z}+a_4}\,,
\end{equation}
where $R_0=22.0\,{\rm Gpc^3\,yr^{-1}}$ denotes the local normalization of the merger rate density and we use the best fit values for the parameters $a_1=6.6\times10^3, a_2=1.6, a_3=2.1$ and $a_4=30$.

We sample a source redshift from the observer frame redshift distribution
\begin{equation}
    P(z_{\rm s}) \propto \frac{R(z_{\rm s})}{1+z_{\rm s}} \frac{d\chi}{dz_{\rm s}} \chi^2\,.
    \label{eq:bbh_zs}
\end{equation}
where $\chi$ denotes the comoving distance to a redshift $z_{\rm s}$. For every source redshift, we randomly sample the binary black hole masses with a probability distribution $P(m_1, m_2)$ given by a Power law + peak model of \citet[][]{Abbott_2021}. We use values for these parameters which are consistent with the analysis of population of detected GW events by the LIGO-VIRGO collaboration, the high mass power-law index $\alpha = 2.63$, the mass ratio power-law index $\beta_q = 1.26$, the low-mass tapering parameter $\delta_m = 4.82 M_\odot$ and the minimum and maximum masses $m_{\rm min} = 4.59 M_\odot$ and $m_{\rm max}=86.22 M_\odot$. For a fraction $\lambda_{\rm peak}=0.10$, we assume that the distribution follows a Gaussian distribution with mean located at $\mu_m=33.07 M_\odot$ with a width $\sigma_{\rm m}=5.69 M_\odot$. We assume our black hole binary systems to be spinless and adopt the IMRPhenomPv2 model \citep{Khan:2016a, Khan:2016b}. 


\subsection{Lens galaxy population}

For a given source, we draw a putative lens galaxy which follows a density profile given by a singular isothermal ellipsoid \citep[e.g.,][]{Kormann1994, Koopmans2009}. We sample the velocity dispersion of this putative lens galaxy  by sampling it from the velocity distribution function of elliptical galaxies given by
\begin{equation}
    P(\sigma)\,d\sigma = \frac{\beta}{\Gamma(\alpha/\beta)}\left(\frac{\sigma}{\sigma_*}\right)^\alpha \exp\left(-\left[\frac{\sigma}{\sigma_*}\right]^{\beta}\right)\frac{d\sigma}{\sigma}
\end{equation}
where we use the values of $\alpha=2.32, \beta=2.67, \sigma_{*} = 161 {\rm \kms}$ \citep{Choi2007}.

We sample the axis ratio, $b$, from the distribution of these ratios of elliptical galaxies from SDSS \citep[see fig. 4 in ][]{padilla2008}. The ellipticity, $q$, of the lens is given by
\begin{equation}
    q = 1- b/a\,.
\end{equation}
The lens redshift is drawn uniformly in the comoving volume between $z=0$ and the source redshift $z_{\rm s}$. Given this lens and source configuration, we compute the probability that such a galaxy at redshift $z_{\rm l}$ could lens the BBH source according to the optical depth,
\begin{equation}
\tau = \frac{\Omega_{\rm lens}}{4\pi}
\end{equation}
where $\Omega_{\rm lens}$ corresponds to the solid angle within which the galaxy should lie in order to cause multiple images of the source. For an SIE, we compute this area enclosed by the caustics numerically as a function of the ellipticity of the system and a fiducial Einstein angle \citep{Keeton2000}. For a given lens velocity dispersion and redshifts of the lens and the source, we use the fact that this solid angle scales with the Einstein angle to calculate $\Omega_{\rm lens}$ for any combination.

We consider the candidate source galaxy as gravitationally lensed by the galaxy if a random deviate between zero and one is smaller than the above optical depth. Given that the lensing optical depth is quite low, we enhance this optical depth by a constant boost factor. We ensure that this boost factor never results in an optical depth larger than unity. This boost factor is equivalent to considering each lens source pair as representative of a larger number of lens source pairs with similar properties. The boost factor helps us to obtain catalog of lens-source pairs efficiently, which have the same distribution of properties as expected of the true lensed events in gravitational wave data.

\subsection{Properties of lensed and unlensed BBH event pairs}

For each lens-source pair, we distribute the position of the source randomly within the source plane such that they form multiply lensed images. We use the software {\sc glafic} \citep[][]{Oguri2010} to solve the lensing equations for the SIE model. Given a random position in the multiply imaged region, we obtain the magnification and the time delay for each of the images of the source. We use the magnification factor to scale the unlensed SNR to a lensed SNR. We consider cases where the second bright lensed counterpart image has a network SNR greater than 8. Note that here, we are ignoring the effect of the antenna patterns of the GW detectors which will change for the two events, which could cause further changes to the network SNR and the selection. These are expected to be secondary effects that affect the distribution of observed properties of lensed events.








For the statistical properties of unlensed BBH event pairs, we randomly draw BBH sources as given in Section~\ref{sec:bbh_src}, and assign them redshifts drawn from the probability distribution in Eq.~\ref{eq:bbh_zs}. We take any pair of these events as candidate lensed event pairs, and compute the probability distribution of their time delays and potential magnification ratios. The apparent magnification ratios of the two candidate lensed event pairs from the unlensed population are computed as the square of the ratios of their inferred distances. The apparent time delays are calculated using the difference of the time of arrival of each of these events. We consider an observation run of 6 months and thus only consider those pairs of events which are separated by this time difference.

\begin{figure}
	\includegraphics[width=\columnwidth]{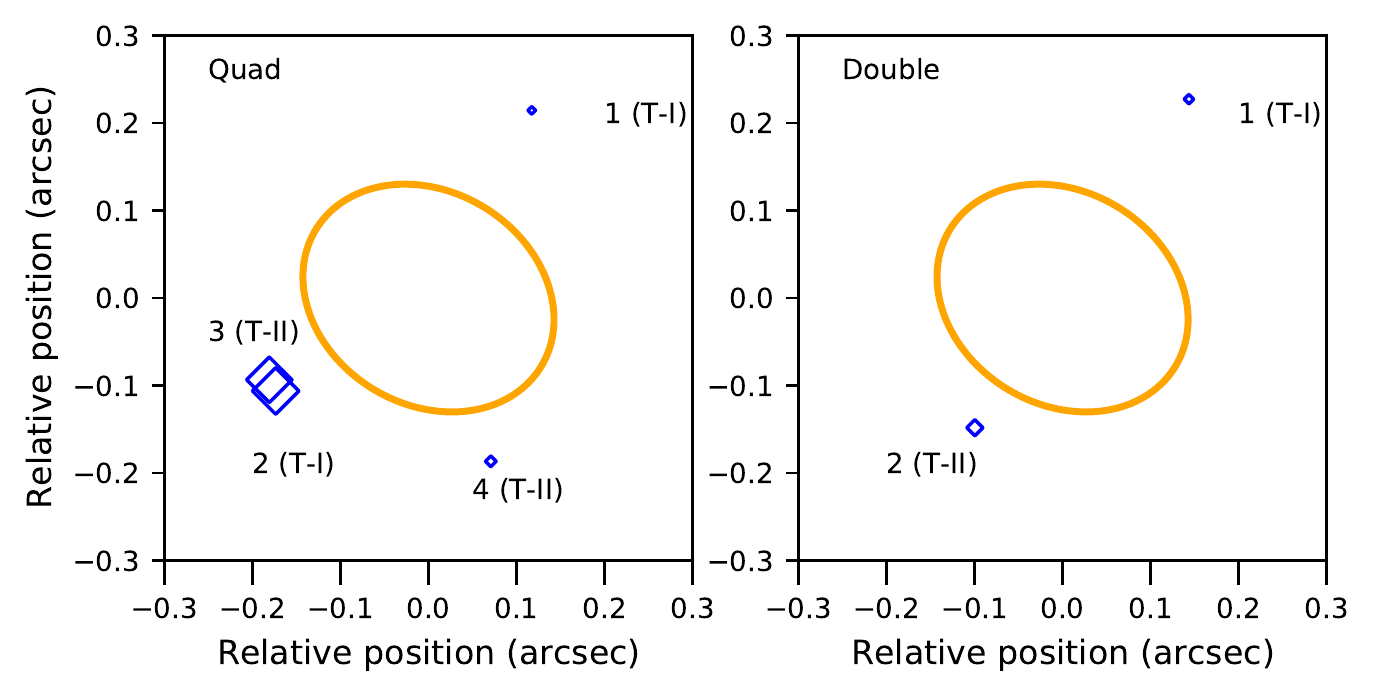}
    \caption{Typical quadruply lensed (quad, left) and doubly lensed
(double,right) configurations shown for illustration. The types (I, II)
 of images are annotated along with the labels 1 to 4 reflecting the
order of arrival of the lensed images. The image positions (diamonds)
are scaled by their lensing magnification and lensing galaxy's
ellipticity and position angle is denoted with the orange circle. }
    \label{fig:illus}
\end{figure}

\begin{figure}
	\includegraphics[scale=0.65]{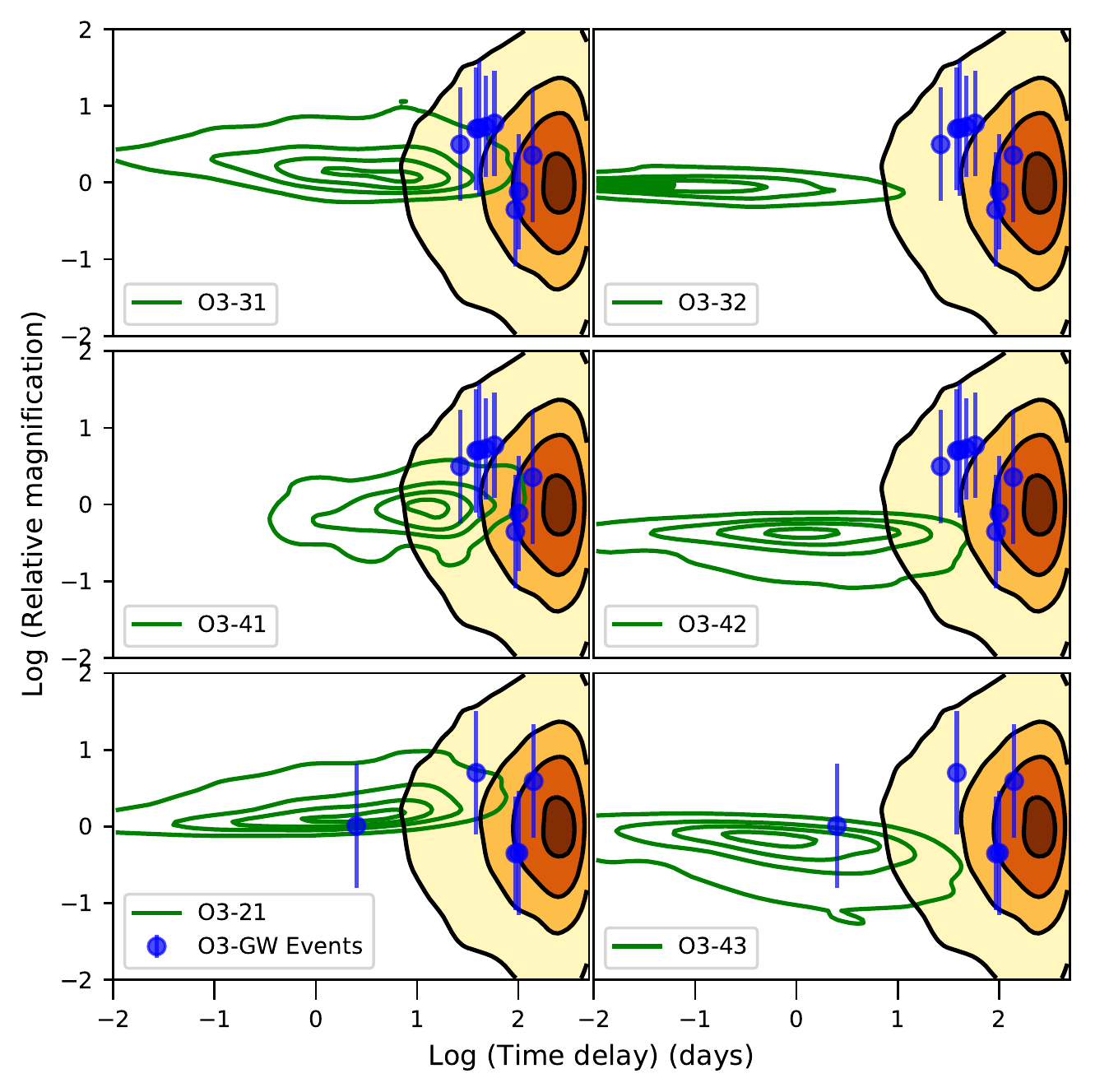} 
    \caption{Time delay and magnification distributions for the simulated lensed (green contours) and unlensed (colormap) population, corresponding to the sensitivity and duration of  LIGO--Virgo's O3a observing run, are shown. Each panel shows a different event pair combination for quads based on the image types. The real candidate lensed GW events from Table 3 of \citet[][]{Abbott2021_lens} with a $\Delta\phi=0$ (bottom row) and $\Delta\phi=\pi/2$ (top two rows) are shown with blue diamonds for comparison. Almost all of the candidate lensed GW events lie on the outskirts of the contours corresponding to simulated lensed event pairs except one (see \sref{sec:statprop} for details). }
    \label{fig:o3cont}
\end{figure}

\section{Statistics of lensed image properties}
\label{sec:statprop}
We have produced a sample of simulated lensed GW systems in double and quad configurations. As we have distributed the sources randomly within the source plane which results in multiple images, the simulated sample has realistic proportions for doubles and quads. We do not account for any observational selection effects owing to the detector duty cycle or detection pipelines currently, however, it should be fairly straightforward to incorporate in our predictions. 

We now explore the lensing magnifications, time delays and image types for each lensed source at $z_s$ provided the second brightest lensed counterpart has a network SNR$>8$, after accounting for magnification effects. We present the joint distributions of the lensing observables for four detector sensitivities, namely, the Advanced LIGO (AL), A+ LIGO (Ap), Cosmic Explorer \citep[CE, e.g.][]{AbbottCE_2017} and Einstein Telescope \citep[ET, e.g.][]{Hild2011, Punturo2014} .

For doubly imaged system (doubles, in short), we expect only two images to be formed and consider them only when both images are above the SNR threshold (see right panel of \fref{fig:illus}). Here, the type I image arrives first followed by the type II image \citep{Schneider1992}. As a result, we expect them to have a lensing phase shift of $\Delta\phi=\pi/2$. 
For a quadruply imaged system (quads, in short), while four images are formed usually, we expect only the brighter two to be detected in most circumstances (see left panel of \fref{fig:illus}). This is why we consider all of the lensed sources which satisfy the condition that the two brightest images are detected. We consider different combinations for the pairs within quads. 
In quads, almost always, the order of arrival is expected to be type I, type I, type II and type II (see \fref{fig:illus}) for the images 1, 2, 3 and 4, respectively \citep{Schneider1992}. If the image pairs are of the same type, type I-type I (denoted as 21\footnote{We use this particular convention to name the pairs. The time delays are defined as $\Delta t_{ij}=t_{i}-t_{j}$ where the $j$-th image is the reference. We list the image which arrives earlier as the reference for measuring time delays.}) or type II-type II (denoted as 43), which correspond to a lensing phase shift $\Delta\phi=0$. If the image pairs are of distinct type, we can have the combinations 31, 41, 32 and 42. 

We note that a type III image may also be formed or be detectable in extremely rare cases. We do not consider this type of images in our analyses as they are usually de-magnified and are highly unlikely to be detected.

We first generate simulated lensed and unlensed pair of events for the sensitivity corresponding to the third observing run (O3) of LIGO--Virgo (see \fref{fig:o3cont}). The green contours in the bottom left panel correspond to the distribution of time delays and relative magnifications for doubles, while those in the rest of the panels correspond to different image combinations in the quads (see legend). The colormap and its associated contours show the distribution of the relative magnifications and time delays between random event pairs analyzed as if they were unlensed.

We can see that the two distributions are quite different and this difference will increase as we consider longer and longer observing runs. This will cause the time delays of the unlensed population to shift further to the right. As we see from \fref{fig:o3cont}, the lensed pairs generally come at short time delays. We also notice that there is some structure to the lensed events in the time delay-magnification plane and these properties are not entirely uncorrelated and are different for the different types of images. For example, shorter time delay events typically have magnification ratios close to unity. Thus, the joint distribution can further help in distinguishing the lensed events from their unlensed counterparts. In the case of doubly imaged lens systems, since both images have to be detected, the magnification ratios are close to unity. For quads, there is a wider variety of distributions of magnifications and time delays.

We also show the promising candidate lensed GW event pairs reported in Table 3 of \citet{Abbott2021_lens} in \fref{fig:o3cont}. We take the 11 pairs with high parameter consistency (i.e. log$_{10}{\mathcal{C}^\mathrm{L}_\mathrm{U}}>4$) which are then divided according to their lensing phase differences and are shown in the respective panels in \fref{fig:o3cont}. Five of the candidate events which have phase differences consistent with the same types of images are shown in the bottom row, while eight of the candidate events have phase differences consistent with similar types of images, and hence are shown  panels. We note that two of the events have high ${\mathcal{C}^\mathrm{L}_\mathrm{U}}$ for both  $\Delta\phi=0$ and $\Delta\phi=3\pi/2$. Most of the candidate GW events mostly overlap with the expected distribution for the unlensed event pairs.

The pair GW190731$-$GW190803 shown in the bottom-left panel (nearly at 0,0) stands out amongst all events to be well separated from the unlensed event pair combinations. If it is truly lensed, this pair is expected to have the same image type and would correspond to a quadruply imaged system. \citet{Abbott2021_lens} quote a value of ${\cal R}^{\rm gal}=8$ for this event pair. We calculate the ${\cal M}^{\rm gal}$ from our distributions in two ways. First, we only consider the brightest lensed pairs (from doubles and quads) from our simulations that are above the SNR threshold and second, we consider only those pairs which have $\Delta\phi=0$ and are above the SNR threshold. The former is similar to the pairs considered in \citet[][]{Haris2018} whereas the latter takes into account the phase difference from \citet[][]{Abbott2021_lens} for the pair GW190731$-$GW190803. Furthermore, in these calculations, we propagate the uncertainties on the inferred luminosity distances in to ${\cal M}$. The mean value for the statistic for the first approach comes out to be ${\cal M}^{\rm gal}=8.5$ consistent with the ${\cal R}^{\rm gal}$ as reported in \citet[][]{Abbott2021_lens}. However when the phases are considered, the mean statistic increases by a factor of $2.5$ such that ${\cal M}^{\rm gal}=22$.

\begin{figure}
	\includegraphics[width=\columnwidth]{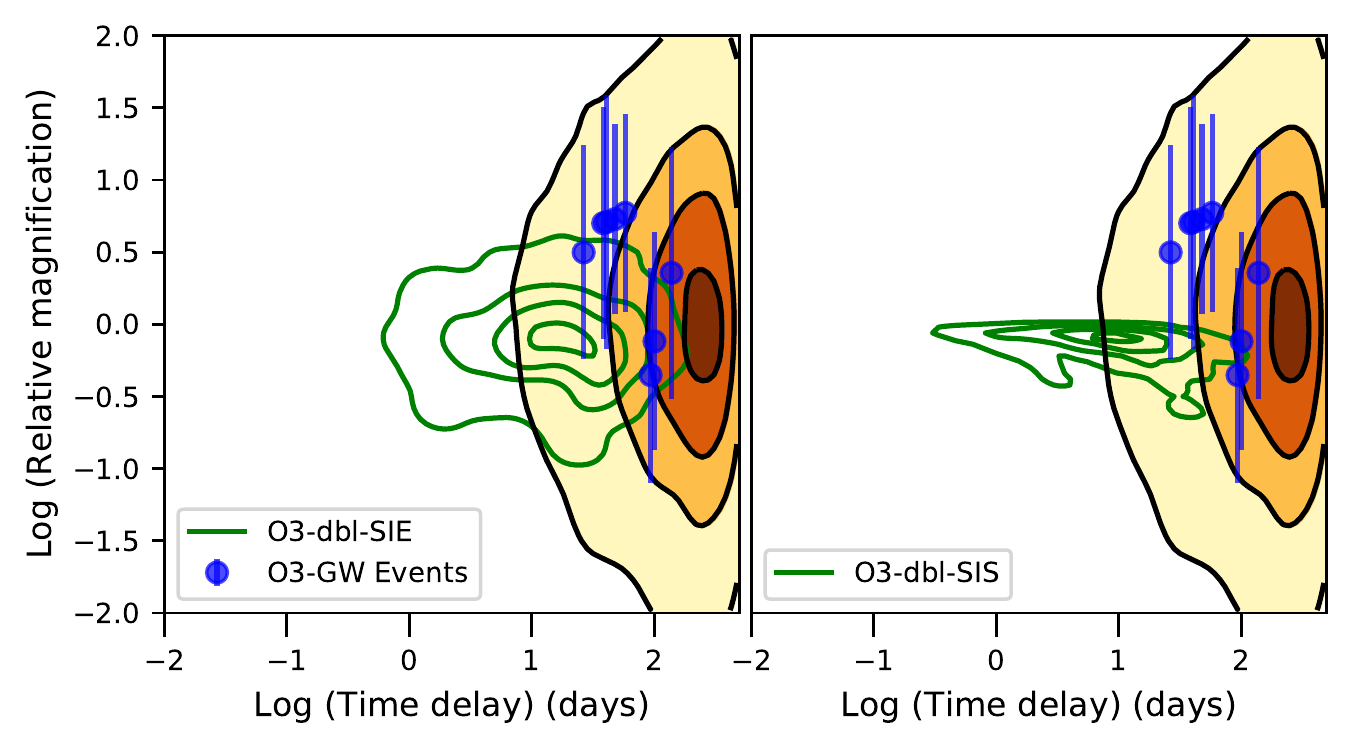} 
    \caption{Time delay and magnification distributions similar to \fref{fig:o3cont} are shown but for doubles where the pair of images will have $\Delta\phi=\pi/2$. The doubles from the SIE models (left) and the SIS models (right) are shown for comparison.}
    \label{fig:o3cont_sis}
\end{figure}

In our analysis, we mainly use the SIE model to describe the lens galaxy potential. The SIE model allows us to explore not only doubles but also quads. We find that the fraction of quads is about 38\% larger than that of doubles. This fraction is consistent with \citet[e.g.][]{Li2018} and also similar to observed fraction of quads from the optical and radio lensing surveys \citep[e.g.][]{Browne2003, Oguri2012}. For the sake of comparison between a relatively simpler mass model of singular isothermal sphere (SIS) with that of the more realistic SIE model for lensing galaxies, we also show the relative magnification and time delays for the doubles produced by SIS in \fref{fig:o3cont_sis}. The time delays of the doubles produced by the SIS and SIE population span the same range (see bottom-left panel of \fref{fig:o3cont}) but the relative magnifications for the SIS tend to occupy a narrower range than with SIE. 

\begin{figure}
	\includegraphics[width=\columnwidth]{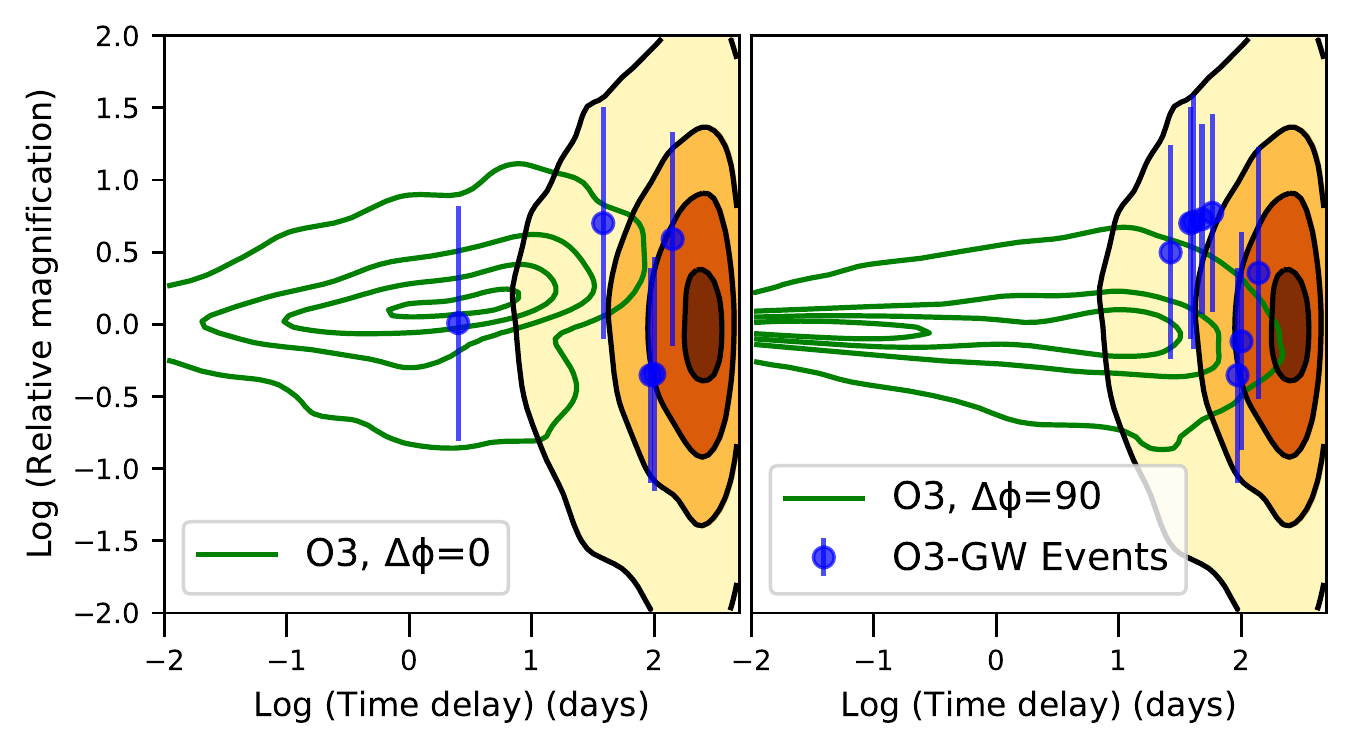}
    \caption{ Time delay and magnification distributions similar to  \fref{fig:o3cont} are shown divided by the phase differences. Images of the same type i.e. $\Delta\phi=0$ are on the left and those of distinct type i.e. $\Delta\phi=\pi/2$ are on the right. Except one, all of the other real GW events are consistent with coming from unlensed population. }
    \label{fig:o3cont_comb}
\end{figure}

The lensing phase difference is an observable property for a pair of events. The left and the right panels of \fref{fig:o3cont_comb} show the joint probability distributions of the time delays and the magnification ratios of event pairs with a phase shift of zero, and those with a phase shift of $\pi/2$, respectively, for our O3 simulations. We also show the O3 events on this joint distributions. 

\begin{figure}
	\includegraphics[width=\columnwidth]{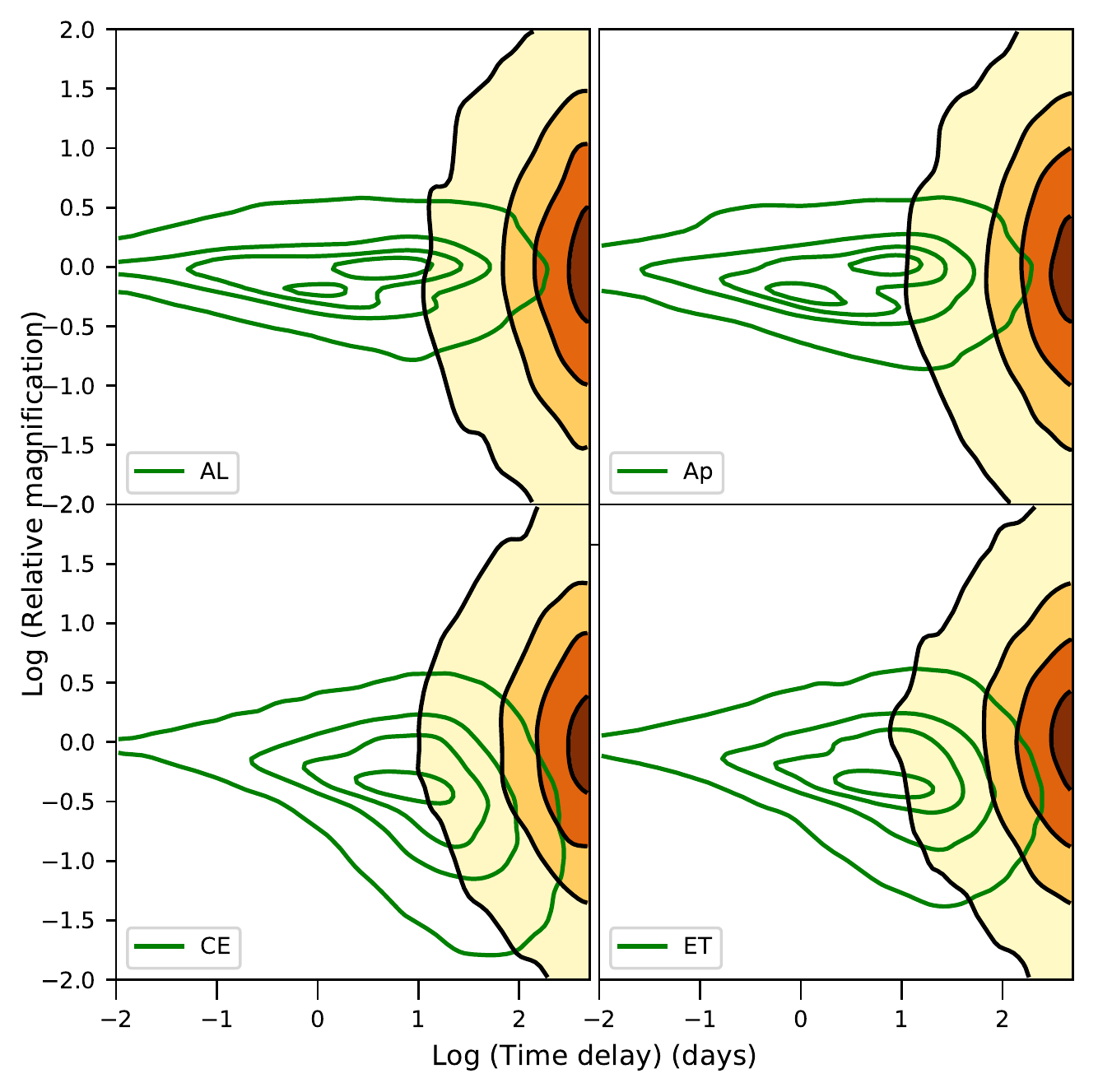}
    \caption{Same as \fref{fig:o3cont} but for Advanced LIGO (AL), A$+$ (Ap), Cosmic Explorer (CE) and Einstein Telescope (ET). 
    Pair of lensed events belong to distinct types i.e. type I - type II (green contours) which include both quads and doubles. Contours with colormap show the distribution for randomly selected unlensed pair of events. }
    \label{fig:phdf90}
\end{figure}

\begin{figure}
	\includegraphics[width=\columnwidth]{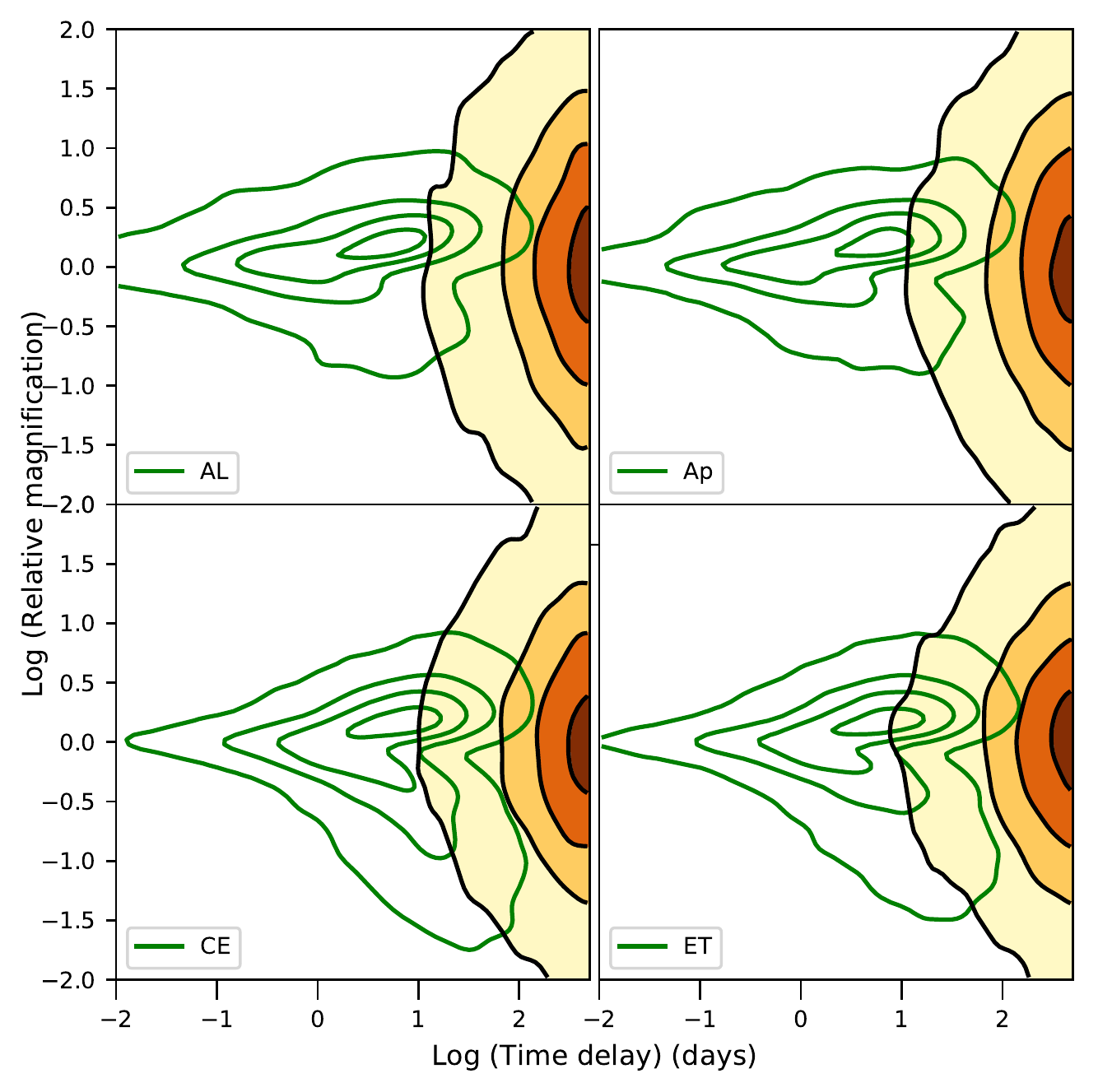}
    \caption{Same as \fref{fig:phdf90} except that the pair of lensed events belong to the same type i.e. type I-type I or type II - type II (green contours). Such pairs will be found only in quads.}
    \label{fig:phdf0}
\end{figure}

Next, we calculate the same probability distributions but for more sensitive detectors such as the Advanced LIGO (AL), A+, CE and ET. Here, we consider events with all possible combinations that could be represented for distinct image types (31, 41, 32 and 42) and the same image type (i.e. 21 and 43) which are shown in \fref{fig:phdf90} and \fref{fig:phdf0}, respectively. As we move to higher and higher sensitivity detectors, many of the fainter lensed counterparts will begin to pass the SNR threshold. Thus, the event pairs with large magnification differences will start to be detected which also tend to have large time delays as these correspond to lensed images located asymmetrically with respect to the lens potential. Our distributions for AL and CE detectors are consistent with those reported in \citet{Oguri2018}.

We also see that the distribution of lensed images of same types are more separated from the unlensed distribution compared to the images of dissimilar types which show a larger overlap in these distributions. However note that the distribution for the unlensed population is entirely driven by the time period during which the event pairs are analyzed. Making the time period of analysis longer will further increase the separation between the lensed and the unlensed population.

%



\section{Summary}
\label{sec:summary}

The discovery for gravitationally lensed events in GW data is both of astrophysical as well as cosmological interest. Multiple GW events with similar intrinsic properties, and overlap in their sky localizations can help establish the lensing hypothesis. However the probability of lensing is low. Therefore, prior probabilities for the properties of gravitational lenses such as their time delays, magnifications and lensing phases, are helpful to distinguish events which happen to have similar intrinsic properties, by chance. In this work, we have developed a framework necessary to compute these prior probabilities. 

We have defined a statistic ${\cal M}_{\rm gal}$ which could be used to improve the previous ranking statistics to identify candidate lensed event pairs. These event pairs could then be analyzed using more accurate but more computationally expensive joint PE analysis codes such as the one presented by \citet{Lo2021}. Our distributions could also be used as priors in the joint PE analysis codes.

To this end, we have carried out simulations of lensed GW events using galaxy scale lenses modelled as SIE for detectors with different sensitivities, both current and those planned in the future. We have computed the image properties of these gravitationally lensed events such as their time delays, magnifications and image types. We have presented the joint probability distributions of these lensing properties for the lensed events and compared it those expected from analyzing random unlensed GW events under the lensing hypothesis. 

Finally, we also compared the properties of candidate lensed GW events from LIGO--Virgo \citep[][]{Abbott2021_lens} with the distributions of the lensing properties from the simulated GW events. In their analysis, the  event pair GW190731$-$GW190803 was reported with the highest ranking statistic ${\cal R}^{\rm gal}=8$. We also find the same event pair to have the highest statistic with a mean value of ${\cal M}^{\rm gal}=22$. We expect that owing to improvements in future detectors, for example, reduced uncertainties in the magnification ratios, our proposed statistic will prove to be much more useful.






\section*{Acknowledgements}
We thank K. Haris, Rico K. L. Lo and P. Ajith for useful discussions. 
The authors are grateful for computational resources provided by the
Inter-University Centre for Astronomy and Astrophysics.

\section*{Data Availability}
Simulated samples of lensed and unlensed populations generated in this paper can be made available upon request to the corresponding author.



\bibliographystyle{mnras}
\bibliography{lensing} 








\bsp	
\label{lastpage}
\end{document}